\documentclass[a4paper,12pt]{article}

\usepackage{amsmath,amssymb,array,float,graphics}
\usepackage{pstricks,pst-node}
\usepackage[figuresleft]{rotating}
\usepackage{rotfloat}
\numberwithin{equation}{section}

\newcommand{\pd}{\partial}

\newcommand{\ket}[1]{\left|#1\right\rangle}

\newcommand{\tr}{\mathop{\mathrm{Tr}}\nolimits}

\newcommand{\Op}{\mathcal{O}}
\newcommand{\ft}[2]{{\textstyle\frac{#1}{#2}}}
\def\tilde{\widetilde}
\def\1bar{1\hskip -.275cm -}
\def\2bar{2\hskip -.275cm -}
\def\3bar{3\hskip -.275cm -}

\newsavebox{\uuunit}

\newcommand{\ads}{AdS$_5\times S^5$ }

\newcommand{\be}{\begin{equation}} \newcommand{\ee}{\end{equation}}
\newcommand{\bea}{\begin{eqnarray}} \newcommand{\eea}{\end{eqnarray}}
\newcommand{\ben}{\begin{displaymath}}
\newcommand{\een}{\end{displaymath}}
 
\newcommand{\nn}{\nonumber}

\newcommand{\nc}{\newcommand}
\nc{\la}{\lambda}
\nc{\alf}{\alpha} \nc{\tht}{\theta}
\nc{\eps}{\epsilon} \nc{\ga}{\gamma} \nc{\Ga}{\Gamma}
\nc{\De}{\Delta} \nc{\de}{\delta} \nc{\si}{\sigma}
\nc{\ka}{\kappa} \nc{\om}{\omega} \nc{\qq}{\quad\quad}
\nc{\nf}{\infty} \nc{\dl}{\mathop{\smash{\cal L}}}
\nc{\ol}{\overline} \nc{\beq}{\begin{equation}}
\nc{\barr}{\begin{array}} \nc{\earr}{\end{array}}
\nc{\eeq}{\end{equation}} \nc{\beqa}{\begin{eqnarray}}
\nc{\dst}{\displaystyle}\nc{\pt}{\partial}
\nc{\eeqa}{\end{eqnarray}} \nc{\nnb}{\nonumber}
\nc{\bs}{\backslash}        \nc{\mbb}{\mathbb}
\nc{\brm}{\begin{remunerate}} \nc{\erm}{\end{remunerate}}
\nc{\vareps}{\varepsilon} \nc{\tb}{\tilde\beta_0} \nc{\ts}{\tilde
s} \nc{\tth}{\tilde \theta}
\newcounter{muni}
\usecounter{muni}
  \nc{\lapdec}{\mathop{\Delta}}

\newenvironment{remunerate}{\begin{list}{{\rm \arabic{muni}.}}
{\usecounter{muni}
\setlength{\leftmargin}{0pt}\setlength{\itemindent}{38pt}}}{\end{list}}
\newdimen\squaresize \squaresize=12pt
\newdimen\thickness \thickness=0.7pt

\def\square#1{\hbox{\vrule width \thickness
   \vbox to \squaresize{\hrule height \thickness\vss
      \hbox to \squaresize{\hss#1\hss}
   \vss\hrule height\thickness}
\unskip\vrule width \thickness} \kern-\thickness}

\def\cut#1{\hbox{\vrule width-1 \thickness
   \vbox to \squaresize{\hrule height-1 \thickness\vss
      \hbox to \squaresize{\hss#1\hss}
   \vss\hrule height-1\thickness}
\unskip\vrule width +4 \thickness} \kern-\thickness}

\def\vsquare#1{\vbox{\square{$#1$}}\kern-\thickness}

\def\young#1{
\vbox{\smallskip\offinterlineskip \halign{&\vsquare{##}\cr #1}}}

\newcommand{\tinyyoung}[1]{
\squaresize=7pt
\thickness=0.4pt
\mbox{\tiny\young{#1}}
\squaresize=12pt
\thickness=0.7pt}

\newcommand{\alg}[1]{\mathfrak{#1}}

\newcommand{\alSU}{\alg{su}}

\newcommand{\alSL}{\alg{sl}}
\newcommand{\alSO}{\alg{so}}

\newcommand{\alPSU}{\alg{psu}}

\nc{\cre}{\color[rgb]{1.00,0.00,0.00}}
\nc{\cgr}{\color[rgb]{0.00,1.00,0.00}}

\makeatletter
\def\hlinewd#1{%
\noalign{\ifnum0=`}\fi\hrule \@height #1 %
\futurelet\reserved@a\@xhline}
\makeatother

\begin{document}

\title{Spin bit models from non-planar ${\cal N}=4$ SYM
}
\author{S. Bellucci, P.-Y. Casteill, J.F. Morales, C. Sochichiu\thanks{On leave from~:
Bogoliubov Lab. Theor. Phys., JINR, 141980 Dubna, Moscow Reg., RUSSIA and
Institutul de Fizic\u a Aplicat\u a A\c S, str. Academiei, nr. 5, Chi\c sin\u au,
MD2028
MOLDOVA.}\\
{\it INFN -- Laboratori Nazionali di Frascati,}\\
{\it Via E. Fermi 40, 00044 Frascati, Italy}}

\maketitle
\begin{abstract}
We study spin models underlying the non-planar dynamics of ${\cal N}=4$ SYM gauge
theory. In particular, we derive the non-local spin chain Hamiltonian
corresponding to the generator of dilatations in the gauge theory at leading order
in $g_{\rm YM}^2 N$ but exact in ${1\over N}$. States in our spin chain-like model
are characterized by a spin-configuration as well as by a linking variable which
describes how sites are connected in the chains. Joining and splitting of
string/traces is implemented by a twist operator acting on the linking variable.
The obtained model is used for the systematic study of non-planar anomalous
dimensions and operator mixing in ${\cal N}=4$ SYM. Beyond other, we identify a
sequence of SYM operators for which corrections to the one-loop anomalous
dimensions stop at the first ${1/N}$ non-planar order.

\end{abstract}

\setlength{\extrarowheight}{2pt}

\section{Introduction}

The correspondence between Yang--Mills and string theories has by
now a long history starting from \cite{'tHooft:1974jz} (see
\cite{'tHooft:2002yn}, for a recent review). The AdS/CFT proposal
\cite{Maldacena:1998re,Gubser:1998bc} for a correspondence between
$\mathcal{N}=4$ super Yang--Mills theory (SYM) and superstring
theory on \ads  is a remarkable realization of these ideas (for a
review see \cite{Aharony:1999ti}). Initially formulated in the
$N\to\infty$ limit, the conjecture in its strong form extends to
finite $N$. It relates the strongly coupled regime of ${\cal N}=4$
SYM to the weakly coupled string theory and viceversa. This property,
which makes out of this correspondence a very strong and efficient
predictive tool, appeared to be an obstacle in proving the duality in
itself.

In \cite{Berenstein:2002jq,Berenstein:2002sa}, Berenstein, Maldacena and Nastase
study the correspondence in the vicinity of some null geodesics of \ads, where the
geometry looks like a gravitational plane wave
\cite{Blau:2001ne}\nocite{Blau:2002dy}-\cite{Blau:2002mw}. On the CFT side this
corresponds to focusing on SYM operators with a large ${\cal R}$-charge. String
theory appears to be solvable \cite{Metsaev:2001bj,Metsaev:2002re} in such a
background and it can be quantitatively compared with predictions coming from
perturbative SYM computations
\cite{Kristjansen:2002bb}\nocite{Gross:2002su,Constable:2002hw,Beisert:2002bb,
Constable:2002vq}-\cite{Spradlin:2003bw} (see
\cite{Plefka:2003nb}\nocite{Spradlin:2003xc}-\cite{Russo:2004kr} for reviews on
the BMN correspondence and references). Later, other limits based on spinning
string solutions were proposed in
\cite{ft1}\nocite{Frolov:2003qc,Frolov:2003tu,Arutyunov:2003uj,%
Beisert:2003ea}-\cite{Beisert:2003xu}. In \cite{Bianchi:2003wx,Beisert:2003te} the
correspondence was tested far from the BMN limit in the free SYM/tensionless
string regime where holography relates the gauge theory to a higher spin gravity
theory on \ads (see
\cite{Sezgin:2002rt}\nocite{Vasiliev:2003ev,Dhar:2003fi,Dhar:2003fi,%
Lindstrom:2003mg,Gopakumar:2003ns,Bonelli:2003zu,Alkalaev:2003qv,%
Sagnotti:2003qa,Metsaev:2003cu,Gopakumar:2004qb}-\cite{Beisert:2004di}
for related works).

At the same time, on the Yang--Mills side considerable progress was
made in the study of the operator mixing and anomalous
dimensions
\cite{Beisert:2002bb},\cite{Beisert:2002ff}\nocite{Beisert:2002tn,Arutyunov:2002rs,%
Bianchi:2000hn,Bianchi:2002rw,Beisert:2003jj}-\cite{Bianchi:2003eg}.
As it was pointed out in \cite{Minahan:2002ve}, the dynamics in
the sector of single-trace bosonic operators of SYM can be mapped
into that of the Heisenberg SO(6) spin one model in such a way that
the matrix of planar one-loop anomalous dimensions is identified
with the Hamiltonian of the spin chain. The Bethe Ansatz
techniques used for diagonalizing the Hamiltonian become then
a powerful tool in determining anomalous dimensions in the gauge
theory. In \cite{Beisert:2003jj,Beisert:2003yb} the result was
generalized to the supersymmetric case.

The spectacular development of the understanding of SYM at large $N$ left somewhat
behind the study of the nonplanar contributions. The latter is expected to
correspond via AdS/CFT to taking into consideration the string production on the
AdS side. String bits
\cite{Verlinde:2002ig}\nocite{Zhou:2002mi}-\cite{Vaman:2002ka} were proposed as a
model which mimics this feature out of (but not very far from) the BMN limit.
Being a reasonably simple and good tool for computing some bosonic quantities, the
string bit model suffers from definite consistency problems related to fermionic
doubling
\cite{Bellucci:2003qi}\nocite{Danielsson:2003yc}\nocite{Bellucci:2003hq}-\cite{Bellucci:2004qi}.

On the SYM side the exact one-loop dilatation operator was derived in
\cite{Beisert:2003jj,Minahan:2002ve,Beisert:2003tq}. When non-planarity is taken
into account, single and multi-trace operators get mixed. In the dual picture this
should correspond to string-string interactions in AdS background which up the
moment is not very well understood.
 Waiting for a better understanding of string physics on
AdS space, one could hope to learn about string interactions there by exploiting
the dual gauge theory picture. This is the main motivation for the present work.
We build a map from the set of multi-trace operators to a model of spins and study
the corresponding spin system which, as we will see, mixes the integrable spin
approach and the string bits one. Such a theory can be called a \emph{spin bit}
model. Since it allows for dynamical splitting and joining of chains and its
variable content is given by spins, the spin bit model differs from both the spin
chain and the string bit models, though being a mixture of both. In particular,
there is no fermion doubling and supersymmetry in the spin bit model is
implemented in a consistent fashion. (In fact, it is inherited from the SYM
theory.)

The spin bit Hamiltonian provides us with a powerful tool simplifying the study of
SYM theory at the non-planar level. The spectrum of anomalous dimensions is
obtained by straightforward diagonalization of the spin bit Hamiltonian. In the
present work we apply this technique to a
 systematic study of anomalous dimensions in ${\cal N}=4$ SYM.
  The results perfectly match those found by quantum field
  theory methods \cite{Anselmi:1998bh}-\cite{Dolan:2001tt} and provide
  a compact and unifying description of these computations.
  This confirms the validity of the expression for the dilatation operator proposed in \cite{Beisert:2003tq}.
  Beyond this, we identify a new sequence of eigenstates starting
  at $\Delta_0=8$ with anomalous
dimensions given by a finite $1/N$ expansion (in fact having the order $(1/N)^1$).

In the limit $N\to \infty$ the spin Hamiltonian becomes the one of ordinary spin
chain and is local and integrable.
 The Hamiltonian and the generator of the total spin are the first
 two charges, in the tower of commuting ones, predicted by
integrability \cite{Beisert:2003tq,Arutyunov:2003rg}. Higher charges are given in
terms of higher powers of next-to-nearest spin generators summed up over the
chain. Corrections in ${1/N}$ spoil locality and integrability. The Hamiltonian
and its higher spin analogs can still be defined in terms of powers of spin
generators but now the next-nearest character is lost and corresponding charges
are no longer commuting among themselves. They can be thought of as broken
symmetries of the would-be integrable system. It would be nice to understand the
role of these broken charges in the theory near the ``integrable" point $N\to
\infty$.

Our paper is organized as follows. In section \ref{sham} we introduce the spin
chain/gauge theory dictionary.  To the exact one-loop dilatation operator of
${\cal N}=4$ SYM we associate the Hamiltonian in the corresponding spin chain. In
section \ref{sanom} we apply the result to the study of non-planar anomalous
dimensions. We determine the exact one-loop characteristic polynomials for the
first few operators sitting in the $\alSU(2)$ and $\alSL(2)$ closed subsectors of
${\cal N}=4$ SYM. Finally in section \ref{discus} we draw some conclusions.
Appendices collect some background material and tables that complement the
discussion in the text.

\section{Spin bits}
\label{sham} In \cite{Minahan:2002ve} Minahan-Zarembo have shown
that planar one-loop corrections to anomalous dimensions of purely
scalar operators in ${\cal N}=4$ SYM can be effectively computed
in terms of an integrable system, the SO(6) spin one model. They
proposed to use Bethe Ansatz for its solution (see
\cite{Faddeev:1996iy} for a review on Bethe Ansatz). In
\cite{Beisert:2003jj} the results were extended to the
supersymmetric case in terms of a $\alPSU(2,2|4)$ integrable spin
chain. Combining the $\alPSU(2,2|4)$ symmetry with the one-loop
planar result, an expression for the non-planar one-loop dilation
operator was finally derived in \cite{Beisert:2003jj}.

In what follows we construct a map from multi-trace SYM operators to
$\alPSU(2,2|4)$ spin states and rewrite the dilation operator as a
 Hamiltonian acting on ordered sets of
these spin states. This action has a non next-nearest character due to joining and
splitting of the chains.

\subsection{Preliminaries}

In this subsection we review the type of quantities we are dealing with as well as
construct the map to spin bit states.

We consider ${\cal N}=4$ SYM theory with the following field content~:
$F_{\mu\nu}$, $\phi^i$, $\lambda^A_\alpha$ and $\bar{\lambda}_{A\dot{\alpha}}$
which are, respectively, the gauge field, six scalars and gauginos\footnote{Here
$i=1,\ldots 6,~A=1,\ldots 4,~\mu=0,\ldots 3,~\alpha,\dot{\alpha}=1,2$. The
abbreviations $\nabla^s \phi,\nabla^s F,\ldots$ stand for $\nabla_{\mu_1}..
\nabla_{\mu_s} \phi^i,\nabla_{\mu_1}.. \nabla_{\mu_s} F_{\mu\nu},\ldots $}. We are
interested in the description of gauge invariant (polynomial) multi-trace
operators in this model. It is convenient to adopt a ``philological'' terminology.
The above fields, as well as their covariant derivatives, form gauge-covariant
``letters'' $W_{A}$ of the SYM ``alphabet"
\begin{equation}
  W_A=\{ \nabla^s \phi,\nabla^s F,\nabla^s \lambda ,\nabla^s
  \bar{\lambda} \}.
\end{equation}
The components of $W_A$ transform in the so called ``singleton" (infinite
dimensional) representation $V_F$ of the ${\cal N}=4$ superconformal algebra (SCA)
$ \alPSU(2,2|4)$. All elementary fields and their derivatives can be obtained by
acting with generators of the SCA on the primary fields $\phi^i$ with $i=1,\ldots
6$ running in the vector representation of $\alSO(6)$. Out of the letters $W_{A}$
one can build gauge invariant ``words'' (single-trace operators) which are traces
of a sequence of $W_{A}$, and out of ``words'' one can produce ``sentences'' which
are sequences of ``words'' (multi-trace operators). For instance, out of $\phi^i$
we can build the word $\tr \phi^{i_1}\ldots \phi^{i_n}$ and the sentences $\tr
\phi^{i_1}\ldots \phi^{i_{n_1}}\tr \phi^{j_1}\ldots \phi^{j_{n_2}}\ldots$.

To each SYM operator (sentence) of length $L$ we can associate a state in a spin
chain or set of chains of the same total length, with symmetry group
$\alPSU(2,2|4)$ and spin states pointed to directions ${A_k}$ in $V_F$ (label $k$
numbers the sites). A state/operator is specified by a spin sequence
$\ket{A_1,\ldots, A_L }$ and by an element $\gamma$ of the $S_L$ permutation group
 \bea
\gamma\equiv (\gamma_1\,
\gamma_2\ldots \gamma_L)~: \quad \left(
\begin{array}{cccc}  a_1  & a_2 &\ldots & a_L\cr
a_{\gamma_1} & a_{\gamma_2} & \ldots  & a_{\gamma_L}\end{array} \right)
 \eea
describing the way different sites in the chain are connected to each other.
Precisely, using these data a  generic multi-trace operator can be written as
\begin{equation}\label{ksentence}
\begin{array}{rl}
 \ket{A_1,\ldots, A_L~;~\gamma} \leftrightarrow & W_{A_1}^{a_1 a_{\gamma_1}}  W_{A_2}^{a_2
  a_{\gamma_2}}\ldots W_{A_L}^{a_L a_{\gamma_L}}\\
  =  & \tr\left( W_{A_1}\dots W_{A_{L_1}} \right)\tr\left(W_{A_{L_1+1}}\dots W_{A_{L_1+L_2}}\right)\dots
  \tr \left( W_{A_{L-L_k+1}}\dots  W_{A_{L}}\right).
\end{array}
\end{equation}
Here $\gamma=({\bf L_1})({\bf L_2})\dots({\bf L_k})$ is a permutation made of
smaller cyclic permutations of $\bf L_m$ elements.
Generically, the permutation group splits in equivalence classes labelled by
$L_1,\dots,L_k$, $\sum L_r=L$ of permutations consisting of cycles of respective
lengths.

The correspondence between operators and spin states is one-to-one, up to
covariant relabelling of the indices. (Cyclic symmetry of the traces is a
particular case of this relabelling.) In fact, different choices of $\gamma\in
S_L$ give different operators \eqref{ksentence}, modulo the equivalence relation
\begin{equation}\label{equiv}
\ket{A_{\sigma_1},\ldots, A_{\sigma_L}~;~\sigma\cdot\gamma\cdot \sigma^{-1}}\sim
\ket{A_1,\ldots, A_L~;~\gamma} =\ket{A_1,\ldots, A_L}\otimes_{S_L} \ket{\gamma}
\end{equation}
where $\otimes_{S_L}$ is the tensor product, modulo the action of
$\sigma\in S_L$\footnote{Here and below products in the
permutation group are understood as $\gamma \cdot \sigma= \gamma
\cdot (\sigma_1\sigma_2\ldots \sigma_L)
=(\sigma_{\gamma_1}\sigma_{\gamma_2}\ldots \sigma_{\gamma_L})$.}.
In particular, when $\sigma=\gamma$ the equivalence \eqref{equiv}
reflects the   cyclicity of the trace
\begin{equation}\label{tr-gamma}
 \ket{A_1,\ldots, A_L~;~\gamma} \sim
 \ket{A_{\gamma_1},\ldots, A_{\gamma_L}~;~\gamma},
\end{equation}

In what follows, we do not restrict ourselves to the canonical form of the
permutation, where $\gamma=({\bf L_1})({\bf L_2})\dots({\bf L_k})$ sends each label
to the immediate next one modulo cyclicity. However, the conjugation \eqref{equiv} can
be used to always rearrange the labels in such a form.

\vspace{1cm}

The correspondence between SYM operators and spin bits allows one to map the
dilatation operator into an operator acting on the spin space. This operator can be
identified with the spin bit Hamiltonian.

In perturbation theory the anomalous dimensions of gauge invariant
operators in ${\cal N}=4$ SYM can be written as follows~:
 \begin{equation}\label{dil0}
 \Delta(g_{\rm YM})=\sum_k H_k \lambda^{k\over 2},
 \end{equation}
with $\lambda={g_{YM}^2 N \over 16\pi^2 }$  being the 't Hooft
coupling. The coefficients in this expansion are given in terms of
effective vertices, i.e. the operators $H_k$. They are determined
by an explicit evaluation of the divergencies of two-point
function Feynman amplitudes. In particular, when an operator $\Op$
renormalizes multiplicatively, the operator \eqref{dil0} becomes
diagonal,
\begin{equation}
  \Delta_{\Op}=\Delta_0+{g_{\rm YM}^2 N\over \pi^2}\, \gamma_{\Op}+\ldots
  ,
\end{equation}
where $\gamma_{\Op}$ is the one-loop anomalous dimension and dots
stand for higher loop corrections.

The tree-level dimensions are the naive ones while one-loop vertices have been
derived in \cite{Beisert:2003jj}
\begin{eqnarray}
 H_0 &=& \Delta_{0A} \, \tr W_A \check{W}^A, \nn\\
 H_2 &=& -{2\over N}\,
 \sum_{j=0}^\infty \,h(j)\,(P_j)_{CD}^{AB} \, : \tr [ W_A, \check{W}^C][ W_B, \check{W}^D]:,
\label{dil}
\end{eqnarray}
where
 \begin{equation}\label{checks}
  \check{W}_{ab}^A=\frac{\pd}{\pd W^{ba}_A}
\end{equation}
and the colons $::$ denote the fact that the derivatives $\check{W}_{ab}^A$ never
act on the letters from the same group inside the colons. $\Delta_{0A}$ are the
classical dimensions of the elementary SYM fields,
 $\Delta_0=1$ for scalar fields $\phi^i$ and derivatives,
$\Delta_0={3\over 2}$ for gauginos and $\Delta_0=2$ for
$F_{\mu\nu}$.
 $(P_j)_{CD}^{AB}$ is the
$\alPSU(2,2|4)$ projector to the irreducible module $V_j$ appearing in the
expansion tensor product of two singletons $V_F$
\begin{equation}
 V_F\times V_F=\sum_{j=0}^\infty \, V_j.
\end{equation}
The first modules $V_0,V_1,V_2$ contain the symmetric, antisymmetric and trace
components in the tensor product of two SYM scalars and their superpartners.
Higher modules $V_j$ contain spin $j-2$ currents and their supersymmetric
completions. Finally $h(j)$ is the harmonic number
$$ h(j)=\sum_{s=1}^j\, {1\over s}.$$
 The effective vertices (\ref{dil}) are manifestly
$\alPSU(2,2|4)$ covariant.  Higher loop contributions involve increasing number of
derivatives and inserted letters.

 \subsection{The Hamiltonian}

Here we derive the one-loop Hamiltonian $H_2$ in the spin chain variables. In
order to do this, we apply the operator $H_2$ given in (\ref{dil}) to the
multi-trace operator corresponding to the spin bit state $\ket{A_1,\ldots, A_L
~;~\gamma}$. As $H_2$ is a second order differential operator one should apply
multiply the Leibnitz rule. Thus, the result will be represented as a sum,
\begin{equation}
  H_2\ket{A_1,\ldots, A_L
  ~;~\gamma}=\sum_{k,l} H_{2,kl}\ket{A_1,\dots,A_k,\dots,A_l,\dots A_L
  ~;~\gamma},
\end{equation}
where $H_{2,kl}$ is the restriction of $H_2$ to the sites with
numbers $k$ and $l$ only.

The two main types of terms emerging from application of (\ref{dil})\footnote{One
can use the so called fusion and fission formulas from \cite{Beisert:2003tq} ~:
$\tr A\,\check{W}_CB\,W_D=\delta_{CD}\tr A\tr B$ and $\tr A\,\check{W}_C\tr
W_DB=\delta_{CD}\tr AB$, where $A$ and $B$ are supposed not to depend on $W$'s.}~:

\begin{eqnarray}\label{ww}
\begin{array}{rl}
\tr (W_A
 \rnode{C}{\check{W}}^{C} W_B \rnode{D}{
 \check{W}}^D)(.. \rnode{k}{W}_{A_k}^{a_k a_{\gamma_k}}.. \rnode{l}{W}_{A_l}^{a_l
 a_{\gamma_l}}..)
 \ncbar[linewidth=.01,nodesep=2pt,arm=.3,angle=-90]{->}{C}{k}
 \ncbar[linewidth=.01,nodesep=2pt,arm=.15,angle=-90]{->}{D}{l}&=\delta^C_{A_k}\delta^D_{A_l} (
 .. W_{A}^{a_la_{\gamma_k}}.. W_{B}^{a_ka_{\gamma_l}}.. )\\[2mm]
&=
 \delta^C_{A_k}\delta^D_{A_l}\ket{A_1 .. A .. B.. A_L
 ~;~\gamma\cdot\sigma_{k l}},\\[2mm]
 \tr (W_A
 \rnode{C}{\check{W}}^{C}\rnode{D}{
 \check{W}}^D W_B )(.. \rnode{k}{W}_{A_k}^{a_k\, a_{\gamma_k}} ..
 \rnode{l}{W}_{A_l}^{a_l\,
 a_{\gamma_l}}..)
 \ncbar[linewidth=.01,nodesep=2pt,arm=.3,angle=-90]{->}{C}{k}
 \ncbar[linewidth=.01,nodesep=2pt,arm=.15,angle=-90]{->}{D}{l}&=\delta^C_{A_k}\delta^D_{A_l}\delta_{a_k\,a_{\gamma_l}} (
 .. W_{A}^{b\,a_{\gamma_k}}.. W_{B}^{a_l\,b}.. )\\[2mm]
&=\delta^C_{A_k}\delta^D_{A_l}\ket{A_1 .. A .. B.. A_L
 ~;~\gamma\cdot\sigma_{k \gamma_l}},
\end{array}
\end{eqnarray}

Here $\sigma_{kl}$ denotes the pairwise permutation of the $k^{th}$ and $l^{th}$
elements, whereas \linebreak $\ket{\{A_1\, ..\, A\, ..\, B\dots A_L
 \}}$ corresponds to the replacement of $W_{A_k}$
and $W_{A_l}$ by $W_A$ and $W_B$ respectively.

The other terms of the Hamiltonian can be obtained from \eqref{ww} by the exchange
$A\leftrightarrow B$ in the first equation and by the simultaneous exchanges
$(A\leftrightarrow B,C\leftrightarrow D)$ in the second one. Using the equivalence
(\ref{equiv}) each of these two terms can be rewritten in the following form~:
\begin{align}\label{identity1}
\ket{\{A_1..B..A..A_L
 \}~;~\gamma\cdot \sigma_{kl}}=\ket{\{A_1..A..B..A_L
 \}~;~\gamma\cdot\sigma_{\gamma_k \gamma_l}}
\end{align}
In order to do this one has to use the property of permutation
$\sigma_{kl}\gamma=\gamma\sigma_{\gamma_k \gamma_l}$ to push the $\sigma$'s to the
right of $\gamma$. For the second term one also needs to relabel the summation
indices $k\leftrightarrow l$.

\begin{figure}[t]
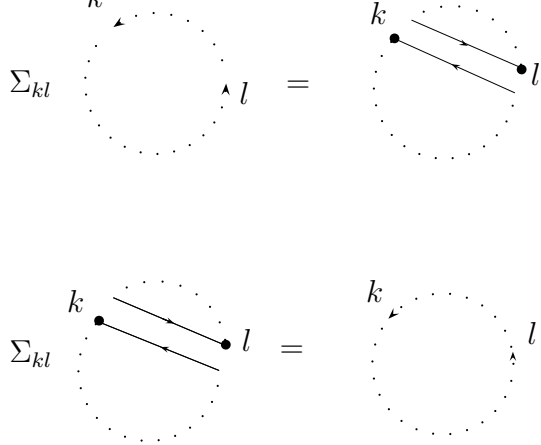

  \centering{\input{1in2a.pic}\\
  \input{2in1a.pic}}
  \caption{Splitting and joining of chains by $\Sigma_{kl}$.}\label{fig:1in2}
\end{figure}

The four terms can be written in a compact form by introducing
the ``two-site Hamiltonian" $H_{k l}$ and the ``twist" operator
$\Sigma_{k l}$ acting on the spin and linking spaces respectively
\begin{eqnarray}\label{ham}
 H_{k l}\,\ket{\{A_1\dots A_L
 \}}&=&4\,\sum_j\, h(j) \,(P_j)_{A_k A_l}^{A B}\,
 \ket{\{A_1\dots A\dots B\dots A_L\}},\\ \label{sigma}
 \Sigma_{k l}\ket{\gamma}&=&
 \begin{cases}
 \ket{\gamma\,\sigma_{kl}}&\text{if }k \neq l
 \\
 N\ket{\gamma},& k=l.
 \end{cases}
\end{eqnarray}
Here $\Sigma_{kl}$ acts as a chain splitting and joining operator as illustrated
in fig. \ref{fig:1in2}. the factor $N$ in the case $k=l$ in eq. \eqref{sigma}
appears because splitting a trace at the same place leads to a chain of length
zero, whose corresponding trace is ${\tr}\, 1=N$. It is important to note that the
operators $\Sigma_{kl}$ act only on the linking variable, while $H_{kl}$ act on
the spin space leaving the link variable unchanged. Therefore, the action of
$H_{kl}$ commutes with those of $\Sigma_{mn}$.

Summing up all ingredients, the one-loop dilation operator acquires the
following form~:
\begin{equation}\label{dil2}
H_2=\ft1{2N} \sum_{k\neq l} H_{k l} \left(\Sigma_{\gamma_k l}+\Sigma_{k \gamma_l}-
\Sigma_{k l}-\Sigma_{\gamma_k \gamma_l} \right).
\end{equation}
Alternatively, using the canonical form for $\gamma=(L_1)...(L_m):
k_i\mapsto [k_i+1]\equiv k_i+1~{\rm mod}~L_i$, with $i=1,2,\ldots, m$
and $k_i$ running inside the $i^{\rm th}$ trace,
eq. (\ref{dil2}) can be rewritten as
\begin{equation}\label{dil3}
H_2=\ft1{2N} \sum_{k \neq l } H_{k  l } \left(
\Sigma_{[k+1], l}+\Sigma_{k,[l+1]}-\Sigma_{k
l}-\Sigma_{[k+1], [l+1]} \right).
\end{equation}
Planar contributions come from terms involving $\Sigma_{k k}$ in
(\ref{dil2},\ref{dil3}), i.e. $l=\gamma_k$ or $k=\gamma_l$~
\begin{equation}\label{h-pl}
H_{2,{\rm planar}}=\sum_{k} H_{k \gamma_{k}}=\sum_{k} H_{k
[k+1]}.
\end{equation}
Summarizing, the Hamiltonian (\ref{dil2}) describes the exact one-loop anomalous
dimensions matrix. As compared to ordinary spin chain description it contains a
new dynamical variable described by a $L$-permutation group element $\gamma$. This
``degree of freedom'' describes\ the chain structure of the configuration and
becomes trivial in the planar case.

There is a certain similarity between our model and string bits
\cite{Verlinde:2002ig,Zhou:2002mi,Vaman:2002ka} in what concerns splitting and
joining of the chains/strings. Notice, however, that our Hamiltonian has a quite
different look from the string bit model Hamiltonian. In contrast to the latter
case, the values of the fields are taken in spin space, rather than a standard
target space, which is the case for string bits. In particular, formulation of the
fermionic sector is completely different. The spin bit model possesses an explicit
supersymmetry, inherited directly from the super Yang-Mills model. Hence there is
no doubling problem, since it would not be compatible with supersymmetry at the
level of the spectrum. The reader may be puzzled by the absence of such phenomena,
which make the supersymmetric string bit model inconsistent. However, as we showed
in \cite{Bellucci:2003hq}, relaxing the requirement that fermions form a
worldsheet spinor structure allows one to formulate a self-consistent
supersymmetric string-like discrete model with no doubling, which is probably the
case in the present spin bit Hamiltonian.

\section{Anomalous dimensions}\label{sanom}

In this section we apply the exact one-loop Hamiltonian
(\ref{dil2}) to the systematic study of non-planar corrections to
anomalous dimensions and mixing for composite operators in ${\cal
N}=4$ SYM.

Our results are in perfect agreement with previous computations
performed via Feynman diagrams \cite{Dolan:2001tt,Ryzhov:2001bp}
and higher spin techniques \cite{Anselmi:1998bh,Anselmi:1998ms}.
Here, anomalous dimensions are derived by diagonalization of the
dilatation operator represented by the spin bit Hamiltonian
(\ref{dil2}). This gives a compact and unified description of the
previous results in the literature and extends easily to higher
scaling dimension states. Once applied to SYM states, the
Hamiltonian (\ref{dil2}) is represented by a block-diagonal matrix
which is easily diagonalizable e.g. by use of a computer.

We will focus on the closed $\alSU(2)$ and $\alSL(2)$
subsectors of the full supersymmetric group $\alPSU(2,2|4)$. The
generalization to the ${\cal N}=4$ supersymmetric spin chain is
straight although technically more involved and will be briefly
described in section \ref{ssusy}.

 We will display exact (one-loop) characteristic polynomials
 for the first few (in general multi-trace)
 operators belonging to the $\alSU(2)$ and $\alSL(2)$ closed sectors of ${\cal N}=4$
 SYM. The anomalous dimensions are the zeroes of such polynomials. In some
 particular cases, where characteristic polynomials nicely factorize,
 an analytic form for the exact one-loop anomalous dimensions will be produced.

\subsection{$\alSU(2)$ spin chain}
\label{ssu2}

The $\alSU(2)$ spin chain can be defined by first restricting to $\alSO(6)$ purely
scalar operators and then choosing a $\alSU(2)$ subgroup inside $\alSO(6)$. In the
case of $\alSO(6)$, there are three irreducible representations appearing in the
expansion of the product of two spin one modules into irreducible components
\cite{Beisert:2003jj}
\begin{subequations}\label{so6}
\begin{align}
  (P_0)^{pq}_{mn} &= \ft12(\delta^{p}_m \delta^{q}_n+ \delta^{q}_m
  \delta^{p}_n)
  - \ft16 \delta_{mn} \delta^{pq},\quad &h(0)=0,\nn\\
  (P_1)^{pq}_{mn} &= \ft12 (\delta^{p}_m \delta^{q}_n-\delta^{q}_m \delta^{p}_n),  \quad &h(1)=1,\nn\\
  (P_2)^{pq}_{mn} &=   \ft16 \delta_{mn} \delta^{pq},
  &h(2)=\ft32.\nn
\end{align}
\end{subequations}
Plugging
this into the two-site Hamiltonian \eqref{ham}, one gets
\begin{equation}\label{h-pair-so6}
  H_{kl}^{\alSO(6)}=2-2\,P_{kl}+K_{kl},
\end{equation}
where $K_{kl}$ and $P_{kl}$ are respectively the trace and the
permutation operators between the $k^{\rm th}$ and $l^{\rm th}$
sites. Restricting to the planar level (\ref{h-pl}), the
Hamiltonian of the integrable $\alSO(6)$ spin chain is found
\cite{Minahan:2002ve}.

Let us focus on the $\alSU(2)$ subsector, i.e. the $\alSU(2)_{j={1\over 2}}$ spin
chain. This
 sector is spanned by
holomorphic operators made out of only two complex scalars, let us say $\phi_0$
and $\phi_1$, transforming in the fundamental representation of $\alSU(2)\in
\alSO(6)$. This corresponds to restricting to SYM states with $\alSU(4)\sim
\alSO(6)$ Dynkin labels $[n,\Delta_0-2n,n]$, with $\Delta_0$ denoting the naive
conformal dimension (i.e. the number of letters) and $n$ being a positive integer
(representing the number of impurities, let us say $\phi_1$ in the $\alSU(2)$
highest weight states). Then, the $\alSU(2)$ spin is identified with the middle
Dynkin label
$$
j=\ft12 \Delta_0-n.
$$
The highest weight states $[n,\Delta_0-2n,n]$ saturate at $g_{\rm YM}=0$  the BPS
like unitarity bounds and sit in $\ft12$-  and $\ft14$- BPS multiplets of the
${\cal N}=4$  SCA for $n=0$ and $n\geq 1$ respectively. When interactions are
turned on $g_{\rm YM}\neq 0$, anomalous dimensions are generated and the bounds
are no longer satisfied. Unprotected $\ft14$-BPS multiplets come together with
semishort multiplets (sharing their dimension)
 to build long multiplets of the superconformal algebra.
Protected $\ft14$-BPS multiplets were
 studied in \cite{D'Hoker:2003vf} (see also \cite{Heslop:2003xu}).
 Here we present a systematic description of anomalous
 dimensions and  mixing for all (in general multitrace)
  $\ft14$-BPS,
  up to $\Delta_0<10$.

The two-site Hamiltonian follows from (\ref{h-pair-so6}) by simply
omitting the trace contribution
\begin{equation}
 H_{kl}=2-2\, P_{kl}\label{ham2}.
\end{equation}
 The exact Hamiltonian is then obtained by plugging (\ref{ham2}) into
 (\ref{dil2}). Using (\ref{identity1}) in terms of operators, \emph {i.e.}  $P_{kl}
 \Sigma_{kl}=\Sigma_{\gamma_k\gamma_l}$, one finds  the $\alSU(2)$
 Hamiltonian :
\begin{equation}\label{Hamsu2}
H_2=\ft1{N} \sum_{k\neq l} (2-2\, P_{kl}) \Sigma_{k \gamma_l}
\end{equation}

 SYM states in the $\alSU(2)$ sector are given by sequences of
 traces made out of $\phi_0$ and $\phi_1$. Since the Hamiltonian (\ref{dil2}) act only
 by either permuting the fields $\phi_0$ and $\phi_1$, or by joining/splitting
 traces, then operators with different numbers of $\phi_0$,$\phi_1$
 do not mix. The states can therefore be characterized by the conformal dimension $\Delta_0$,
 which is equal to the total number of fields, and by $n$, i.e.
 the number of impurities $\phi_1$. The anomalous dimension matrix can be found by first
 listing all inequivalent multi-trace operators for a fixed $(\Delta_0,n)$ and then acting upon them with the Hamiltonian.

\vspace*{1cm}
 The $\alSU(2)$ symmetry can be used to reduce the entropy of the analysis.
 Indeed SYM states are organized in irreducible representations
 of $\alSU(2)$, with all components in the same $\alSU(2)$ multiplet
 sharing the same anomalous dimension. This is clearly the case since the
 Hamiltonian commutes with the operator of total spin. We can then focus on
 $\alSU(2)$ highest weight states (h.w.s.).
 Irreducible representations of
 $\alSU(2)$ are specified by Young tableaux with at most two rows.
 Therefore, we will use Young tableaux in order to represent our
 operators (see Appendix for details).

For example, the tableau $Y=\tinyyoung{1&2 \cr 3&4 \cr}$ will stand for the
following operator\footnote{ In our convention, the symmetrization $(~)$ and
antisymmetrization
$\rnode{A}{\phantom{a}}~~\rnode{B}{\phantom{a}}\ncbar[linewidth=.01,nodesep=2pt,arm=.07,angle=-90]{-}{A}{B}$
symbols do not take into account the usual ${1\over p!}$
 factor~: $T_{(kl)}=T_{kl}+T_{lk}$ and
$T_{\rnode{A}{\scriptstyle k}\rnode{B}{\scriptstyle
l}}=T_{kl}-T_{lk}\ncbar[linewidth=.01,nodesep=2pt,arm=.07,angle=-90]{-}{A}{B}$.\vspace*{2mm}}~:

$$\young{1&2 \cr 3&4 \cr}\raisebox{5pt}{$\dst~~ ={1\over |Y| }\,\tr\phi_{(\rnode{A}{\scriptstyle i_1}}
\phi_{\rnode{C}{\scriptstyle i_2})} \phi_{(\rnode{B}{\scriptstyle
i_3}} \phi_{\rnode{D} {\scriptstyle i_4})}
\ncbar[linewidth=.01,nodesep=2pt,arm=.2,angle=-90]{-}{A}{B}
\ncbar[linewidth=.01,nodesep=2pt,arm=.1,angle=-90]{-}{C}{D}$}~.$$
 Here the tensor $\tr\phi_{i_1} \phi_{i_2} \phi_{i_3} \phi_{i_4}$ is
projected according to the operator
 $\tinyyoung{1&2 \cr 3&4
\cr}$~ by first symmetrizing along indices in the same row
($i_1\leftrightarrow i_2,i_3\leftrightarrow i_4$) and then
antisymmetrizing indices along the columns ($i_1\leftrightarrow
i_3,i_2\leftrightarrow i_4$). Notice that the two actions do not
commute and therefore the resulting operator is no longer
symmetric along the rows.
 $|Y|$ is a combinatorial factor (see formula (\ref{su2y}) in Appendix {\bf B}).
 Multi-trace operators will be represented in the Young tableaux
by thick columns indicating where new traces start. Finally, we
will not write the numbers in the boxes of the Young tableaux
filled in the natural order, e.g. $\tinyyoung{&& \cr &
\cr}\equiv\tinyyoung{1&2&3 \cr 4&5 \cr}$~. The Appendix {\bf B}
give all details on the construction of such Young tableaux
states.

\vspace*{1cm}

As an illustration, let us consider an example of the sector of operators with
$(\Delta,n)=(4,2)$.  The list of considered states is
$$\left\{\begin{array}{ll}
\tr\phi_0^2\phi_1^2 &=\left| 0011 ;(2341)\right>\\
\tr(\phi_0\phi_1)^2 &=\left| 0101 ;(2341)\right>\\
\tr\phi_0^2\tr\phi_1^2 &=\left| 0011 ;(2143)\right>\\
(\tr\phi_0\phi_1)^2&=\left| 0101 ;(2143)\right>~.
\end{array}\right.$$
This sector is closed under the action of the Hamiltonian given by
(\ref{Hamsu2}),
 with indices $k$ and $l$ running from 1 to 4.
 Let us consider first the double-trace state $\tr\phi_0^2\tr\phi_1^2$.
When the Hamiltonian acts on this state  one has to compute
expressions as
$$\begin{array}{lll}
(1-P_{32})\Sigma_{3 \gamma_2 }| 0011; (2143) \rangle
&=(1-P_{23})| 0011  \rangle&\otimes~\Sigma_{1 3}|(2143) \rangle\\
&=\big(|0011\rangle -|0101 \rangle)&\otimes~  |(2143) \sigma_{13}\rangle\\
&=\big(|0011 \rangle -|0101\rangle)&\otimes~  |(2341) \rangle\\
&\multicolumn{2}{l}{=|0011 ;(2341)\rangle -|0101 ;(2341)\rangle}\\
&\multicolumn{2}{l}{=\tr\phi_0^2 \phi_1^2-\tr (\phi_0\phi_1)^2~.}
\end{array}$$
 All various terms in (\ref{dil2}) give
similar contributions. After summing them up, one ends with
$$H_2 \tr\phi_0^2 \tr\phi_1^2=\ft{16}{N}\, \tr\phi_0^2 \phi_1^2-\ft{16}{N}\,\tr (\phi_0\phi_1)^2~.$$
 This kind of computation can easily be implemented
with computer software as \emph{Mathe\-matica}.

 In the same way, one finds
$$\left\{\begin{array}{ll}
H_2\tr\phi_0^2\phi_1^2&=4\tr\phi_0^2\phi_1^2-4\tr(\phi_0\phi_1)^2\\
H_2\tr(\phi_0\phi_1)^2 &=-8\tr\phi_0^2\phi_1^2+8\tr(\phi_0\phi_1)^2\\
H_2(\tr\phi_0\phi_1)^2&=-\frac{8}{N}\tr\phi_0^2\phi_1^2+\frac{8}{N}\tr(\phi_0\phi_1)^2~.
\end{array}\right.$$
The anomalous dimensions $\gamma$ are then given by
$\hbox{Det}(\gamma-\frac{H_2}{16})$.
  As a result one finds eigenvalue $\gamma=0$ (triple degenerate) and (non-degenerate) $\gamma=3/4$.
 Two out of the three $\gamma=0$ states sit in the completely symmetric
 $j=2$ $\alSU(2)$ multiplets with highest weight state\footnote{Here highest weight states correspond to Young tableaux $Y$ with $\Delta_0-n$
boxes (filled with $\phi_0$'s) in the first row and $n$ boxes
(filled with $\phi_1$) in the second row}~:
\bea
 j=2 &&  \tinyyoung{&&& \cr}\, {}_{\rm hws}= \tr \phi_0^4  \nn\\
   && \tinyyoung{&\cut{}&& \cr}\, {}_{\rm hws}= \tr \phi_0^2 \tr \phi_0^2\label{d4}
\eea

     The remaining two eigenstates are $\alSU(2)$ singlets ($j=0$)
 can be written as\footnote{Here and below we assume N large
 enough ($N>\Delta_0$), in order to avoid non trivial identifications
 between  single and multi-trace operators. The generalization to
 small values of $N$ is straightforward. For instance, taking $N=2$ in (\ref{d4}), the
 first $j=0$ state vanishes identically, while the two $j=0$
 states are related to each other and should be counted only once.}
 \bea
 j=0 && \tinyyoung{&\cut{} \cr & \cr}\, {}_{\rm hws}-\frac{4}{N} \tinyyoung{& \cr & \cr}\, {}_{\rm hws}=
 \ft23 \, \tr \phi_0^2 \tr \phi_1^2 - \ft23 (\tr \phi_0 \phi_1)^2  -\ft{4}{3 N} \tr \phi_0
[ \phi_0, \phi_1] \phi_1 \nn \\
  && \tinyyoung{& \cr & \cr}\, {}_{\rm hws}=  \ft13 \tr \phi_0
[ \phi_0, \phi_1] \phi_1. \nn \eea with
 $\gamma=0$ and $\gamma=3/4$ respectively.  The latter
 corresponds to the $[2,0,2]$ scalar in the long Konishi multiplet.

  In a similar way one proceeds for a different number of impurities $n$.
  The one row tableau $\tinyyoung{&&& \cr}$ and
$\tinyyoung{&\cut{}&& \cr}$ collect all completely symmetric
combinations with $n=0,1,\ldots, 4$ impurities with h.w.s. $n=0$
given by (\ref{d4}). There are $2\times 5$ states of this type.
Altogether they build two spin $j=2$ $\alSU(2)$ multiplets. The
anomalous dimension Hamiltonian acts trivially on these states
since they are symmetric and therefore $\Delta=\Delta_0$.

Analogously, in the case of higher scaling dimensions, one first lists $\alSU(2)$
highest weight states for a given $\Delta_0$ and then diagonalizes the spin bit
Hamiltonian (\ref{dil2},\ref{dil3}) in the corresponding subspace.
We collect  in table \ref{table:t1su2} the characteristic
polynomials
 for the first few $\alSU(2)$ states up to $\Delta_0=7$.
 The exact anomalous dimensions are the zeroes of these
polynomials.
 In particular, the order of the polynomial gives the number of
 different (in general multi-trace) operators
 built out of $n$ $\phi_0$'s and $\Delta_0-n$ $\phi_1$'s.
 Finally we display in the last column the corresponding eigenvalues at leading
 order in $N$ (planar anomalous dimensions).
In table \ref{table:t3su2}, we collect the exact
 eigenstates and anomalous dimensions up to $\Delta_0=6$.
  Notice that although anomalous
dimensions for states in this table do not receive ${1\over N}$
corrections, a non-trivial mixing between single and multi-trace
operators is at work.

\begin{table}[h]
$$\begin{array}{|l|l|l|l|}
\hline
\Delta _{0} & $n$ & \gamma_{\rm exact} & \gamma_{\rm planar} \\[2mm]
\hline
4 & 2 & \left ( -\frac{3}{4}+x\right )  \, x &  0,\frac{3}{4}  \\[2mm]
\hline
5 & 2 & \left ( -\frac{1}{2}+x\right )  \, x &  0,\frac{1}{2}  \\[2mm]
\hline 6 & 2 & x^{3} \left ( \ft{10}{N^2}-15  -\ft{40}{N^2} \, x+80 \,
\, x-128   \, x^{2}
  +64   \, x^{3}\right )  &  0^{3},\frac{3}{4},\frac{5\pm \sqrt{5}}{8} \\[2mm]
\hline
  & 3 & \left ( -\frac{3}{4}+x\right )  &
  \frac{3}{4},  \\[2mm]
\hline 7 & 2 & x^{4} \left ( 9 \, +\ft{42}{N^2} \, x-78   \,
x-\ft{80}{N^2} \, x^{2}+232
   \, x^{2}-288   \, x^{3}+128   \, x^{4}\right )  &  0^{4},
   \frac{1}{4},\frac{1}{2},\left ( \frac{3}{4}\right ) ^{2}  \\[2mm]
\hline
  & 3 & {\begin{array}{l} \left ( -\frac{5}{8}+x\right )  \, x^{2} \,
  \left (-\frac{9+\sqrt{1+\frac{160}{N^2}}}{16} \,
     +x\right )  \, \left (-\frac{9-\sqrt{1+\ft{160}{N^2}}}{16} \,
     +x\right )
       \end{array}}
  &  {\begin{array}{l} 0^{2},\frac{1}{2},\left ( \frac{5}{8}\right ) ^{2}
      \end{array}}\\[2mm]
\hline
\end{array}$$
\caption{Characteristic polynomials for $\alSU(2)$ h.w.s. with
$\Delta\leq 7$.\label{table:t1su2}}
\end{table}

We display only $\alSU(2)$ highest weight states. More precisely a
state in the table at $(\Delta_0,n)$ is the highest weight state
of a spin $j=\ft12\Delta_0-n$ representation of
$\alSU(2)$\footnote{$\alSU(2)$ descendants are given by
$\ket{i_1,\ldots i_L}\to \sum_k \,\delta_{i_k,0}\,\ket{i_1,\ldots
i_L}|_{i_k\to 1}$}.
  This implies that each characteristic
polynomial for a state $(\Delta_0,n)$ in the table
 will be replicated at $\sum_{m=0}^{\Delta-2n}(\Delta_0,n+m)$.
For example the polynomial at $(\Delta_0,n)=(6,2)$ appears again at $(6,3)$ and $(6,4)$.
Altogether they form a spin $j=1$ $\alSU(2)$ multiplet.
These redundant states are not displayed.

 In addition we omit multi-trace operators with $n=0,1$ number of impurities
 and their $\alSU(2)$ descendants. They lead
 to states with exact conformal dimensions $\Delta_0$ protected
 from both loops and non-planar corrections. States with $n=0$ sit in
 the vacuum multiplets with highest weight states ${\rm Tr} \, \phi_0^\Delta$.
 They correspond to all possible cuts
 of the chiral vacuum state  $\tr \phi_0^{\Delta_0}$. Similarly states
 with $n=1$ arise from cuttings of  $\tr \phi_1 \phi_0^{\Delta_0-1}$.
They correspond to the highest weight states of $\alSU(2)$ irreducible representations
with spin $j=\ft12\Delta_0$
and $j=\ft12(\Delta_0-1)$ respectively. Generating functions for states with $n=0,1$
can be written as\footnote{The omission of the term $J=1$ in the product
corresponds to the fact that $\tr \phi_0=0$ in $SU(N)$.}~:
\bea
n=0 &&\quad\quad \prod_{J=2}^\infty {1\over 1-\tr \phi_0^{J}}\nn\\
n=1 &&\quad\quad \tr\left[ {\phi_0 \phi_1 \over 1-\phi_0}\right]
 \prod_{J=2}^\infty {1\over 1-\tr \phi_0^{J}.}\label{mult01}
\eea
 For instance, at $(\Delta,n)=(4,1)$, we find two ``one impurity"
 states: $\tr \phi_0^3 \phi_1$ and $\tr \phi_0 \phi_1\, \tr \phi_0^2$.
 The full characteristic polynomial at $(\Delta,n)=(4,2)$ is then given
 by the product of the polynomial in the table with $x^2$ coming from the
 $\alSU(2)$ descendants of the two $n=1$ states.

 The most interesting features start showing up at $n=2$. Non-planar corrections first appear
 for $\Delta_0=4$ where a single trace can split into two double-trace
 operators, each made out of two letters.
  The $(\Delta_0,n)=(6,2),(7,2)$ cases were studied
 in \cite{Beisert:2003tq,Bianchi:2003eg}.
 The one-loop anomalous dimensions (the zeros of the characteristic polynomials)
 in these cases can be written as an infinite ${1\over N}$-expansion.

 In particular circumstances, the non-planar characteristic polynomials
nicely factorize and an explicit form for the exact anomalous
dimensions to all orders in ${1\over N}$ can be written.
Table \ref{table:t2su2} in the appendix collects some relevant examples
where exact (to all order in $1/N$) expressions for one-loop anomalous dimensions can be written.
The first case is the pair of states at $\Delta_0=7$ with $n=3$
impurities
%
 where the infinite series of ${1\over N}$
corrections reconstruct a square root \cite{Ryzhov:2001bp}.
 Even more interesting is a (presumably infinite) series of
 operators starting with $(\Delta_0,n)=(8,3)$, $(9,4)$, $(10,3)$, \dots
whose one-loop anomalous dimensions get non-planar corrections
only at the first order in ${1\over N}$
\bea
\Delta^{(8,3)}_{\pm}&=& 8+{g_{YM}^2 N\over \pi^2}\, \left(\frac34 \pm \frac3{4N}\right)\nn\\
\Delta^{(9,4)}_{\pm}&=& 9+{g_{YM}^2 N\over \pi^2}\, \left(\frac58 \pm \frac3{4N}\right)\nn\\
\Delta^{(10,3)}_{\pm}&=& 10+{g_{YM}^2 N\over \pi^2}\,
\left(\frac34 \pm \frac3{2N}\right) \label{exact} \eea The
presence of $1/N$ corrections rather than the more familiar
$1/N^2$ may appear surprising.
Notice however that such corrections come
always in pairs with opposite signs and therefore the
corresponding characteristic polynomials depend only on $1/N^2$.
Non-planar corrections result in a splitting of the degenerate
energy levels at planar order (see \cite{Beisert:2002ff} for
more details).

It is important to stress that non-planar corrections to the one-loop Hamiltonian
are only of order $1/N$ corresponding to the joining-splitting string vertex
$g_s\sim 1/N$. Anomalous dimensions (energy levels) follow from this Hamiltonian
via diagonalization. On the string theory side the eigenstates (\ref{exact})
correspond to bound states which are a mixture of single and multi-string states.

For convenience of the reader, table \ref{table:t2su2} is also given in terms of
traces, rather than with Young tableaux, in table \ref{table:t4su2} in the
appendix.

\subsection{$\alSL(2)$ spin chain}

Now we consider states belonging to a $\alSL(2)$ subgroup. This
subgroup is generated by a single scalar field component, say
$\phi_0$, and all its covariant derivatives along e.g. the
first direction: $\phi^{n}={1\over n!} {\cal D}_1^n\, \phi_0$. A
state in the $\alSL(2)$ sector is then specified by the sequence
$\ket{\{ n_i \}}\equiv \{\phi^{n_1},\phi^{n_2},\ldots
,\phi^{n_L}\}$ and the linking variable $\ket{\gamma}$. Notice
that unlike the $\alSU(2)$ case,
  representations at each site are now infinite dimensional, i.e. $n_k$ runs
from zero to infinity.

The two-site Hamiltonian is given by \cite{Beisert:2003jj}:
\begin{eqnarray}\label{h-pair-sl2}
  H_{kl}\,\phi_k^{m}\phi_l^{n-m}
  =\left[h(m)+h(n-m) \right]\,\phi_k^{m}\phi_l^{n-m}-\sum_{m'=0}^{n} {\delta_{m\neq m'}\over |m-m'|}
  \phi_k^{m'}\phi_l^{n-m'}.\nn
\end{eqnarray}
Plugging in (\ref{dil2}) we can find the exact $\alSL(2)$
Hamiltonian. The spectrum of non-planar characteristic polynomials
for the first few states in this sector is displayed in table
\ref{table:tsl2} (see \cite{Anselmi:1998bh,Anselmi:1998ms} for
previous revsults in this sector). Here again, states are labelled
by the classical dimension $\Delta_0$ and the number of
impurities, i.e. derivatives, $n$. Since each derivative
contributes once to the dimension $\Delta_0$, the number of
letters used in building the states in table \ref{table:tsl2} is
$L=\Delta_0-n$. Once more, we omit states with $n=0,1$ impurities
and $SL(2)$ descendants. Now
 $\alSL(2)$ descendants of each line $(\Delta_0,n)$ span an
infinite
 tower $\sum_{m=0}^\infty (\Delta_0+m,n+m)$ of
eigenstates found by acting with $m$ derivatives on a given
eigenstate\footnote{A derivative corresponds to send
$\ket{n_1,\ldots ,n_L}\to \sum_k (n_k+1) \, \ket{n_1,\ldots
,(n_k+1),\ldots ,n_L}$.}

\begin{table}[h]
$$\begin{array}{|l|l|l|l|}
\hline
\Delta _{0} & \hbox{n} & \gamma_{\rm exact} & \gamma_{\rm planar} \\[2mm]
\hline
4  & 2 & \left ( -\frac{3}{4}+x\right )  \, x &  0,\frac{3}{4}  \\[2mm]
\hline
5 & 2 & \left ( -\frac{1}{2}+x\right )  \, x &  0,\frac{1}{2}  \\[2mm]
\hline 6 & 2 & x^{3} \left ( \ft{10}{N^2}-15  -\ft{40}{N^2} \, x+80   x-128
    \, x^{2}+64  \, x^{3}\right )  &  0^{3},\frac{3}{4},\frac{5\pm \sqrt{5}}{8}  \\[2mm]
  & 3 & \left ( -\frac{15}{16}+x\right ) ^{2}
    &  \left ( \frac{15}{16}\right ) ^{2}  \\[2mm]
   & 4 & \left ( -\frac{25}{24}+x\right )  &
    \frac{25}{24} \\[2mm]
\hline 7 & 2 & x^{4} \left ( 9  +\ft{42}{N^2} \, x-78
\, x-\ft{80}{N^2} \, x^{2}+232
    \, x^{2}-288   \, x^{3}+128   \, x^{4}\right )
   &  0^{4},\frac{1}{4},\frac{1}{2},\left ( \frac{3}{4}\right ) ^{2} \\[2mm]
   & 3 & \left ( -\frac{3}{4}+x\right ) ^{3} \,
   &  \left ( \frac{3}{4}\right ) ^{3}   \\[2mm]
  & 4 &   \left ( -\frac{3}{4}+x\right )
   &  \frac{3}{4}    \\
\hline
\end{array}$$
\caption{Characteristic polynomials for $\alSL(2)$ highest weight states with
$\Delta_0\leq 7$ .\label{table:tsl2}}
\end{table}
 It is instructive to compare the
$\alSU(2)$ and $\alSL(2)$ tables \ref{table:t1su2} and
\ref{table:tsl2}. States in the two tables are often related by
supersymmetry. If this is the case, their full non-planar
characteristic polynomials of anomalous dimensions (not only the
planar contributions) should coincide. Indeed, a simple inspection
shows that the full $n=2$ characteristic polynomials perfectly
match. This is agreement with the results in \cite{Beisert:2002tn} where
$n=2$ impurity states has been shown to belong to the so called
"BMN supermultiplets" with highest weight state primary in the
 $[0,\Delta_0-2,0]$ representation of $SU(4)$.

\subsection{Supersymmetric spin chain}
\label{ssusy}

For completeness we briefly describe next the generalization to the supersymmetric
case. The two-site Hamiltonian for the ${\cal N}=4$ supersymmetric spin chain was
derived in \cite{Beisert:2002tn}, in terms of the so called harmonic action. In
this formalism SYM letters are represented by acting with any number of bosonic
$(a_\alpha, b_{\dot{\alpha}})$ and fermionic oscillators $(c_r,d_{\dot{r}})$,
$\alpha,\dot{\alpha},r,\dot{r}=1,2$ on a Fock space vacuum $a^{n_a} b^{n_b}
c^{n_c} d^{n_d} |0\rangle$  subjected to the condition \be C=n_a-n_b+n_c-n_d=0.
\label{c0}\ee The two-site Hamiltonian reads \cite{Beisert:2003jj} \be H_{kl}
\,|s_1,..s_n\rangle=\sum_{s_i'} c_{n,n_{kl},n_{lk}}\,
\delta_{C_k,0}\delta_{C_l,0}\,|s'_1,..s'_n\rangle. \label{hamsusy} \ee
 Here $n$ is the total number of oscillators and $s_i,s'_i=k,l$ denote their position.
Remarkably the Hamiltonian does not depend on the type of insertion but only
 on the positions of the insertions labelled by the sequences of $s_i$'s. Also,
$\delta_{C_k,0},\delta_{C_l,0}$ ensure that the $C=0$ condition
(\ref{c0}) holds at each site. Finally  the coefficients
$c_{n,n_{kl},n_{lk}}$ are given by\bea
c_{n,n_{kl},n_{lk}}&=&(-)^{1+n_{kl} n_{lk}}\,
{\Gamma(\ft12(n_{kl}+n_{lk}))\Gamma(1-\ft12(n-n_{kl}-n_{lk}))\over
\Gamma(1+\ft12 n)},\nn\\
 c_{n,0,0}&=& h(\ft12 n),\nn
 \eea
with $n_{kl},n_{lk}$ denoting the number of oscillators hopping from the $k$ to the
$l$ site and viceversa.

 The non-planar Hamiltonian is given in terms of (\ref{hamsusy}) via
(\ref{dil2},\ref{dil3}).
 The supersymmetry invariance of $H_2$ follows
from the fact  that $H_{kl}$ commutes with both $\alPSU(2,2|4)$
\cite{Beisert:2003jj} and
$\Sigma_{k,l}$.
  The Hamiltonian $H_2$ determines the full
non-planar corrections to one-loop
anomalous dimensions in ${\cal
N}=4$ SYM theory.


\section{Discussion}\label{discus}
In this paper we apply spin chain techniques to the study of
nonplanar corrections to the one-loop anomalous dimension matrix
for composite operators in ${\cal N}=4$ SYM.
Eigenstates in the spin bit model describe, via AdS/CFT correspondence, bound states
in String Field Theory, where string interactions are weighted by 1/N.

While the planar approximation of this theory leads to the
description in terms of integrable spin chains, taking into
account the nonplanarity leads to the appearance of new degrees of
freedom, i.e. the linking variable, in the spin chain model.
Interactions result in dynamical fissions and fusions of the
chains. This effect resembles the string splitting and joining in
the string field theory. Also the discrete model we obtained has
some similarity with spin network models introduced by Penrose
\cite{Penrose:1971} (for a review see \cite{Rovelli:1995ac}) in
order to describe discrete gravity.

In order to demonstrate the power of the spin bit approach we use it to perform the
computation of anomalous dimensions of composite operators in SYM theories.
Thus, we reproduce a number of already known results and produce new ones, as well.
Out of the total symmetry group $\alPSU(2,2|4)$ of the model we focus on closed
subsectors corresponding to the following subgroups: $\alSU(2)$ and $\alSL(2)$.
We provide a detailed analysis of anomalous dimensions and
mixing in these sectors.
States in the two sectors are not completely independent but are
related in many instances by supersymmetry. When this is the case, the exact anomalous dimensions in
the two sectors match. Anomalous dimensions are encoded in
characteristic polynomials, but an analytic form for the zeros is often hard to be
extracted. Remarkably, in particular circumstances the characteristic polynomials
nicely factorize and an analytic form for the exact anomalous dimensions can be
written. This is the case for a pair of eigenstates at $\Delta_0=7$ with three
impurities, where non-planar corrections in ${1\over N^2}$ are summed up to
reconstruct an exact square root \cite{Ryzhov:2001bp}. Even more surprisingly, we identify a new sequence
of paired operators where the conformal dimensions get corrected only at order
${1\over N}$. The string interpretation of this result
and whether it remains true also to higher loops still
remains to be clarified.

Beyond the application to the study of non-planar corrections in ${\cal N}=4$ SYM
theory, the spin models under consideration here have their own interest as an
example of a polymer model with dynamical splitting and joining
and a nontrivial discrete model with supersymmetry.
Eqs. (\ref{dil2},\ref{dil3}) give a natural extension of any spin model
(integrable or not) to account for decaying and fusions of the chain. This
generalization can be thought as a sort of gauging of the global symmetry $k_i\to
k_i+1$ present in the planar (next-to-nearest) interaction (\ref{h-pl}). Indeed,
this symmetry is enhanced in (\ref{dil2},\ref{dil3}) to $k\to \sigma_k$, with
$\sigma\in S_L$. The gauging is provided by the ``connection" $\Sigma_{kl}$.

In our approach we did not use the whole power of Bethe Ansatz and integrability.
Whether integrable techniques can be used efficiently at least in the framework of
perturbation theory near the integrable point $N\to \infty$ remains to be seen. To
this purpose one should compute the scalar products (formfactors) of Bethe states
with arbitrary spin states. There is some approach in the literature to this
issue\footnote{We thank N. Slavnov for pointing our attention to this research.}
\cite{Korepin:1982gg} (see \cite{NSlavnov} for a review of recent developments).

Another issue we left beyond our consideration is related to the fact that, at
finite $N$, nonperturbative effects in SYM theory begin to take place. So far, we
do not know what will be their effect on our analysis.

\subsection*{Acknowledgements}
We benefited from useful discussions with G.Arutyunov, N.Beisert,
M.Bianchi, S.Di Matteo, C.Natoli, F.Palumbo, N.Slavnov and Y.
Stanev. This work was partially supported by NATO Collaborative
Linkage Grant PST.CLG. 97938, INTAS-00-00254 grant,
INTAS-00-00262, RF Presidential grants MD-252.2003.02,
NS-1252.2003.2, INTAS grant 03-51-6346, RFBR-DFG grant 436 RYS
113/669/0-2, RFBR grant 03-02-16193 and the European Community's
Human Potential Programme under contract HPRN-CT-2000-00131
Quantum Spacetime.

\appendix

\section{Young Tableaux}

In this appendix, we collect some group theory  material and
explain our notations for Young tableaux used in section
\ref{ssu2}. A Young tableaux is a row-decreasing diagram made out
of boxes $\tinyyoung{\cr}$~, representing the fundamental
representation of a group $G$ (here $G=S_m$ or $G=SU(m)$).
 Tensor products $\tinyyoung{\cr}^{\otimes k}$, i.e. tensors $T_{i_1\ldots i_k}$ with
 $i_s=1,2,\ldots m$, decompose into a sum of irreducible
 representations of $G$ characterized by Young tableaux specifying how indices are
 symmetrized or antisymmetrized.

We define the Young symmetrizer
${S_{j_1,j_2,\cdots,j_p}}^{i_1,i_2,\cdots,i_p}$ by\footnote{We
do not put the usual renormalisation factor ${1\over p!}$ for
further simplicity.}~:
$${\bf S}={S_{j_1,j_2,\cdots,j_p}}^{i_1,i_2,\cdots,i_p}=\sum_{\sigma\in S_p}{\delta_{j_1}}^{i_{\sigma_1}}{\delta_{j_2}}^{i_{\sigma_2}}
\cdots{\delta_{j_p}}^{i_{\sigma_p}}$$
and the Young antisymmetrizer ${A_{j_1,j_2,\cdots,j_p}}^{i_1,i_2,\cdots,i_p}$ by~:
$${\bf A}={A_{j_1,j_2,\cdots,j_p}}^{i_1,i_2,\cdots,i_p}=\sum_{\sigma\in S_p}\eps(\sigma){\delta_{j_1}}^{i_{\sigma_1}} {\delta_{j_2}}^{i_{\sigma_2}}
\cdots{\delta_{j_p}}^{i_{\sigma_p}}~.$$

A Young tableau $Y$ can be seen as the projection
\be
T^Y_{i_1,i_2,\ldots i_p}=
{1\over |Y| }({\bf A} {\bf S})_{i_1,i_2,\ldots i_p}^{j_1,j_2,\ldots j_p}
T_{j_1,j_2,\ldots j_p}
\label{proj}
\ee
 with ${\bf S}$(${\bf A}$) denoting the operator that (anti)symmetrizes indices in the same
 row (column): starting
with a tensor $T_{i_1,i_2,\cdots,i_p}$, we first apply
symmetrizers according to the rows of $Y$, \emph{then}
antisymmetrizers according to the columns of $Y$. For example, applying
$Y=\tinyyoung{1&2&3 \cr 4&5 \cr}$ on $T$ means that we first
symmetrize the indices $(i_1,i_2,i_3)$ and $(i_4,i_5)$,
\emph{then} antisymmetrize the \emph{resulting} tensor
$T^S_{i_1 i_2 i_3 i_4 i_5}\equiv T_{(i_1 i_2 i_3)(i_4 i_5)}$ on
indices $[i_1,i_4]$ and $[i_2,i_5]$ \footnote{In practice, this is
equivalent to first antisymmetrize and then symmetrize, but
acting on the \emph{positions} of the indices appearing in the
tensors rather than on their labels.}. Notice that the two actions
do not commute and therefore the resulting tensor is no longer symmetric
on $(i_1,i_2,i_3)$ and $(i_4,i_5)$.

There are two combinatorial factors that characterize a tableau $Y$~: $|Y|$ and $f_{Y}$. For example, we have

$$\begin{array}{rll}
\raisebox{5pt}{$|Y|=$}&\young{5&4&2&1 \cr 2&1 \cr}&
 \raisebox{5pt}{$= 80$} \nn\\
f_{Y}=&
\scalebox{0.5}{
 \setlength{\arrayrulewidth}{4\arrayrulewidth}
\begin{tabular}{@{\vrule width 2pt}c@{\vrule width 2pt}c@{\vrule width 2pt}c@{\vrule width 2pt}c@{\vrule width 2pt}}\hline
m & \,m+1\, & \,m+2\, & \,m+3\, \\ \hline
\,m-1\, & m \\ \cline{1-2}
\end{tabular}}&= (m-1)m^2(m+1)(m+2)(m+3)~.
\end{array}$$

 They depend only on the shape of the tableaux. Coefficient $|Y|$
is given by the following hook formula~: write the sum of the
number of boxes to the bottom and to the right, plus one, inside
each box of the Young tableau, then multiply all these numbers.
The result gives the overall coefficient in (\ref{proj}) that
ensures that the tableau is indeed a projector.
 The two quantities determine the dimension $d_Y$ of a tableau (the number
 of independent components in $T^Y_{i_1,\ldots i_k}$),
 and the multiplicities $n_Y$ of a given tableau shape in
 the tensor product $\tinyyoung{\cr}^{\otimes k}$.
$$
d_Y ={f_Y\over |Y|} \qquad~~~~~~~~~~ n_Y={k!\over |Y|}
 $$
e.g. \bea \tinyyoung{\cr}^{3} &=&
\tinyyoung{&&\cr}+2\,\tinyyoung{&\cr
\cr}+\tinyyoung{\cr\cr\cr}\nn\\
m^3&=& {1\over 3!} m(m+1)(m+2)+2{1\over 3} m(m^2-1)+{1\over 3!}
m(m-1)(m-2) \nn\eea
 Here $n_Y$ are the coefficients in front of the tableaux, $d_Y$
 the dimensions.
 Using these conventions, we are now able to represent the
$\alSU(2)$ eigenstates as Young tableaux by identifying these with
their action on a tensor of the form
$Tr[\phi_{i_1},\phi_{i_2},\cdots,\phi_{i_k}]$, with $i_p=0,1$. As
we are considering $\alSU(2)$, the only Young tableaux we should
use have at maximum two rows. The following rules have been
followed:
\begin{itemize}
\item In cases of multiple traces, thick columns in the Young
tableaux give the positions where new traces start. \item After
the projection, the indices $i_p$ for which $p$ is in the first
$(\Delta_0-n)$ boxes take the value 0, while the  $i_p$ for which
$p$ is in the last $n$ boxes are set to 1.
\end{itemize}

As an example we give, for the case $(\Delta_0=6,n=3)$, the procedure that constructs $\tinyyoung{1&\cut{2}&3&4 \cr 5&6 \cr}$~:
$$ \left\{\begin{array}{ll}
T_{i_1 i_2 i_3 i_4 i_5 i_6}&= \tr\phi_{i_1}\phi_{i_2}\tr\phi_{i_3}\phi_{i_4}\phi_{i_5}\phi_{i_6}\\[2mm]
{T^S}_{i_1 i_2 i_3 i_4 i_5 i_6}&= T_{(i_1 i_2 i_3 i_4)(i_5 i_6)}\\[2mm]
{T^Y}_{i_1 i_2 i_3 i_4 i_5 i_6}&=
{1\over |Y|}
~{T^S}_{\rnode{A}{\scriptstyle i_1}
 \rnode{C}{\scriptstyle i_2} i_3 i_4 \rnode{B}{\scriptstyle i_5}\rnode{D}
{\scriptstyle i_6}}
\ncbar[linewidth=.01,nodesep=2pt,arm=.2,angle=-90]{-}{A}{B}
\ncbar[linewidth=.01,nodesep=2pt,arm=.1,angle=-90]{-}{C}{D}\\[4mm]
\tinyyoung{1&\cut{2}&3&4 \cr 5&6 \cr} &= {T^Y}_{000111}
\end{array}
\right.$$
The result is then
$$
\raisebox{10pt}{$\young{1&\cut{2}&3&4 \cr 5&6 \cr}$}~~
\begin{array}{l}
=\frac{1}{|Y|}~\delta_{0~0~0~1~1~1}^{k_1k_2k_3k_4k_5k_6}~(\delta_{k_3\,k_4}^{j_3\,j_4}~A_{k_1k_5}^{j_1j_5}~A_{k_2k_6}^{j_2j_6}
)~(S_{j_1j_2j_3j_4}^{i_1i_2i_3i_4}~S_{j_5j_6}^{i_5i_6})~\tr\phi_{i_1}\phi_{i_2}\tr\phi_{i_3}\phi_{i_4}\phi_{i_5}\phi_{i_6}\\[2mm]
=\frac{1}{10}\tr{\phi_1}^2\tr{\phi_0}^3\phi_1
+\frac{1}{5}\tr\phi_0\phi_1\tr{\phi_0}^2{\phi_1}^2-\frac25\tr\phi_0\phi_1\tr(\phi_0\phi_1)^2\\[2mm]
\hfill+\frac{1}{10}\tr{\phi_0}^2\tr\phi_0{\phi_1}^3~.
\end{array}
$$
For simplicity, we do not write the numbers in the boxes when these are filled in
the natural order, e.g. $\tinyyoung{&& \cr & \cr}\equiv\tinyyoung{1&2&3 \cr 4&5 \cr}$.

For the $SU(2)$ case, a simple formula gives the
$|Y|$ coefficient:
\be
\left|~ \begin{array}{|c|l}\hline \multicolumn{2}{|c|}{
\longleftarrow ~~~n_1 ~~~\longrightarrow } \\ \hline \leftarrow~
n_2 ~\rightarrow & \\ \cline{1-1}
\end{array}~\right|=\frac{(n_1+1)!~n_2!}{n_1-n_2+1}~.
\label{su2y} \ee
Finally, the spin of such a tableau is given by
$j=\frac12(n_1-n_2)$.

\section{$\alSU(2)$ anomalous dimension eigensystems}

Here we collect three $\alSU(2)$ tables:~ table \ref{table:t3su2} lists all
eigenstates for $\Delta\leq 6$,
table \ref{table:t2su2} in the appendix collects the first few exact eigenstates for
operators with rational $\gamma\neq 0$ anomalous dimensions,
table \ref{table:t4su2} gives the
translations of table \ref{table:t2su2} in the main text.

\begin{table}[H]
$$\begin{array}{|l|l|l|l|}
\hline
\Delta _{0} & $n$ &\hbox{Eigenvectors} & \gamma_{\rm exact} \\[2mm] \hline

4&0& \tinyyoung{&&& \cr} & 0\\[2mm]
 & & \tinyyoung{&\cut{}&& \cr} &\\[2mm]\cline{2-3}
 &2& \tinyyoung{&\cut{} \cr & \cr}-\frac{4}{N} \tinyyoung{& \cr & \cr} &    \\[2mm] \cline{3-4}
 & & \tinyyoung{& \cr & \cr}  & \frac34\\[2mm] \hline\hline

5&0& \tinyyoung{&&&&\cr} & 0\\[2mm]
 & & \tinyyoung{&\cut{}&&& \cr} &\\[2mm]\cline{2-3}
 &1& \tinyyoung{&\cut{}&& \cr \cr} & \\[2mm]\cline{2-3}
 &2& \tinyyoung{&\cut{}& \cr & \cr}-\frac{2}{N} \tinyyoung{&& \cr & \cr} &  \\[2mm] \cline{3-4}
 & & \tinyyoung{&& \cr & \cr}  & \frac12\\[2mm] \hline\hline

6&0& \tinyyoung{&\cut{}&&&& \cr} & 0\\[2mm]
 & & \tinyyoung{&&\cut{}&&& \cr} &\\[2mm]
 & & \tinyyoung{&\cut{}&&\cut{}&& \cr} & \\[2mm]\cline{2-3}
 &1& \tinyyoung{&\cut{}&&& \cr \cr} & \\[2mm]\cline{2-3}
 &2& \tinyyoung{&&&\cut{} \cr & \cr}+\frac1N \tinyyoung{&\cut{}&&\cut{} \cr & \cr}
       -\frac{8}{N} \tinyyoung{&&& \cr & \cr} &  \\[2mm]
 & & \tinyyoung{&\cut{}&&\cut{}\cr & \cr}+\frac8N\tinyyoung{&\cut{}&& \cr &\cr}
       +\frac{16}{3N} \tinyyoung{&&\cut{}& \cr & \cr}+\frac{4}{N} \tinyyoung{&&&\cut{} \cr & \cr} &  \\[2mm]
 & & \tinyyoung{&&\cut{}& \cr & \cr}-\frac38\tinyyoung{&&&\cut{}\cr & \cr}
       +\frac{3}{2N} \tinyyoung{&&& \cr & \cr} &  \\[2mm]\cline{2-4}
 & 3 & \tinyyoung{1&2&4 \cr 3&5&6 \cr} &{3\over 4}\\[2mm]
       \hline
\end{array}$$
\caption{Eigenstates for $\alSU(2)$ states with $\Delta\leq 6$
.\label{table:t3su2}}

\end{table}

\newpage

\begin{table}[h]
$$\begin{array}{|l|l|l|l|}
\hline
\Delta _{0} & $n$ &\hbox{Eigenstates} & \gamma_{\rm exact} \\[2mm] \hline

4&2& \tinyyoung{& \cr &\cr}$$ &  \frac34\\[2mm] \hline

5&2& \tinyyoung{&& \cr &\cr}$$ &  \frac12\\[2mm] \hline

6&3& \tinyyoung{1&2&4 \cr 3&5&6 \cr}$$ &  \frac34\\[2mm] \hline

7&3& \tinyyoung{&&& \cr && \cr}+\frac23\tinyyoung{1&3&4&5 \cr 2&6&7 \cr} & \frac58 \\[2mm] \cline{3-4}
 & & \frac{1}{4}\left(1\pm\sqrt{1+\frac{160}{N^2}}\right)\tinyyoung{&&& \cr && \cr}
 +\frac1N\tinyyoung{&\cut{}&&\cr &&\cr} &
     \frac{1}{16}\left(9\pm\sqrt{1+\frac{160}{N^2}}\right) \\[2mm] \hline

8&3& \left(1 \pm \frac2N \right)\tinyyoung{&&&& \cr && \cr}+\left(1 \pm \frac2N \right)\tinyyoung{1&3&4&5&6 \cr 2&7&8 \cr}
     -\left(1 \pm \frac4N \right)\tinyyoung{1&2&4&5&6 \cr 3&7&8 \cr}

 & \frac34 \pm \frac3{4N} \\[2mm]
 & & \hfill \pm\left(1 \pm \frac3N \right)\left(\tinyyoung{&\cut{}&&& \cr && \cr}\mp\tinyyoung{1&\cut{2}&3&4&7 \cr 5&6&8 \cr}
    +\tinyyoung{1&\cut{3}&4&5&6 \cr 2&7&8 \cr}\right) &  \\[2mm] \hline

9&4& \tinyyoung{&&&& \cr &&& \cr}+\tinyyoung{1&2&3&4&6 \cr 5&7&8&9 \cr}+ \tinyyoung{1&2&3&6&7 \cr 4&5&8&9 \cr}-\tinyyoung{1&2&3&5&7 \cr 4&6&8&9 \cr}
 & \frac58 \pm \frac3{4N} \\[2mm]
 & & \hfill \pm \frac12\tinyyoung{&\cut{}&&& \cr &&& \cr} \pm \tinyyoung{1&\cut{2}&3&4&6 \cr 5&7&8&9 \cr} \mp \frac12\tinyyoung{1&\cut{2}&3&6&7 \cr 4&5&8&9
 \cr}
  \pm \frac12\tinyyoung{1&\cut{2}&3&5&7 \cr 4&6&8&9 \cr}
   &  \\[2mm] \hline


\end{array}$$
\caption{Analytic $\alSU(2)$ eigenstates with $\gamma\neq 0$ and
$\Delta<10$.\label{table:t2su2}}
\end{table}

\newpage

\begin{sidewaystable}[H]
\begin{tabular}{|l|l|l|l|}
\hline
$\Delta _{0}$ & $n$ &\hbox{Eigenvectors} & $\gamma_{\rm exact}$ \\[2mm]
\hline

4&2 & $\frac{1}{3}\, \tr[\phi _{0},\phi _{1}]\phi _{1}\phi _{0}$
 & $\frac34$ \\[2mm] \hline\hline

5&2&  $\frac{1}{2} \, \tr[\phi _{0},\phi _{1}]{\phi _{1}}{\phi
_{0}}^2$
 & $\frac12$\\[2mm] \hline\hline

6&3& $\frac{1}{2} \, \tr[\phi _{0},\phi _{1}]{\phi _{1}}{\phi _{0}}^2{\phi_{1}}$
 &  $\frac34$\\[2mm] \hline\hline

7&3& $\frac{1}{3} \, \tr[\phi _{1},\phi _{0}]{\phi _{1}}{\phi _{0}}^3{\phi
_{1}}$
 & $\frac58$ \\[2mm] \cline{3-4}

 & & $\frac{1}{3N}\,\left(\tr\phi _{0}\phi _{1}
     \, \tr[\phi _{1},\phi _{0}]\phi _{1}{\phi _{0}}^2
     +\tr {\phi _{0}}^2 \, \tr[\phi _{0},\phi _{1}]{\phi
     _{1}}^2\phi_{0}\right)
     -\frac{1}{20}
 \, {\scriptstyle \left ( 1\pm\sqrt{1+\frac{160}{N^{2}}}\right)}
 \, \left ( \tr {\phi _{0}}^3\phi _{1}\phi_{0}{\phi _{1}}^2
 +\tr {\phi _{0}}^3{\phi _{1}}^2\phi _{0}\phi _{1}\right )$~~
  & $\frac{1}{16}{\scriptstyle \left ( 9\pm\sqrt{1+\frac{160}{N^{2}}}\right)}$\\[2mm]

  & & \hfill $+\frac{1}{30} \, {\scriptstyle \left ( 1\pm\sqrt{1+\frac{160}{N^{2}}}\right)}
\, \left ( \tr {\phi _{0}}^4{\phi_{1}}^3
 +\tr ({\phi _{0}}^2\phi _{1})^2\phi _{1}
  +\tr \phi_{0}(\phi _{0}\phi _{1})^3 \right) $ & \\[2mm]\hline\hline

8&3& $\left (\pm\frac12+\frac{3}{2N}\right )  \, \tr{\phi_{0}}^2
\, \tr [\phi _{1},\phi _{0}]\phi _{1}{\phi _{0}}^2\phi_{1}
+\left (\frac12\pm\frac{1}{N}\right )  \, \tr {[\phi _{1},\phi _{0}]\phi _{1}\phi _{0}}^4\phi _{1}
+\left (\frac12\pm\frac{2}{N}\right )  \,  \tr
 [\phi_0,\phi_1]{\phi_{0}}^2{\phi_{1}}^2{\phi_{0}}^2$
& $\frac34 \pm \frac3{4N}$\\[2mm]\hline\hline

9&4& $\frac13\left(\pm\tr \phi _{0}\phi _{1}  \, \tr [\phi _{0},\phi _{1}]\phi_{1}{\phi _{0}}^3\phi _{1}
\pm\tr{\phi _{0}}^2  \, \tr [\phi _{1},\phi_{0}]\phi _{0}{\phi _{1}}^3\phi
_{0}+\tr [\phi _{0},\phi _{1}]{\phi _{0}}^2{\phi _{1}}^3{\phi _{0}}^2\right.$
 & $\frac58\pm\frac{3}{4N}$ \\[2mm]

 & & \hfill$\left.
+\tr [\phi _{0},\phi _{1}]{\phi _{1}}^2\phi_{0}\phi_{1}{\phi _{0}}^3
+\tr [\phi _{0},\phi _{1}]{\phi _{0}}^3\phi _{1},\phi _{0}{\phi _{1}}^2
+ \tr [\phi _{0},\phi _{1}]\phi _{1}{\phi _{0}}^3\phi _{1}\phi _{0}\phi _{1}
+\tr [\phi _{0},\phi _{1}]\phi _{1}\phi _{0}\phi_{1}{\phi _{0}}^3\phi
_{1}\right)$
 &  \\[2mm]\hline\hline

10&3& $\pm\left ( \frac12\pm\frac{9}{2N}+\frac{21}{2N^2}\right )
\, (\tr {\phi _{0}}^2) ^{2} \, \tr [\phi _{0},\phi _{1}]\phi _{1}{\phi _{0}}^2\phi _{1}
+\left ( 1\pm\frac{3}{N} \right )  \, \left ( \frac12\pm\frac2N \right ) \, \tr{\phi _{0}}^4
\, \tr [\phi _{0},\phi _{1}]\phi _{1}{\phi _{0}}^2\phi
_{1}$

 & $\frac34 \pm \frac3{2N}$\\[2mm]

 & & \hfill$+\frac2N  \left ( \pm 1+\frac4N\right )
 \, \tr{\phi _{0}}^3  \,\tr [\phi _{1},\phi _{0}]\phi _{1}{\phi _{0}}^3\phi _{1}
 +\left ( 1\pm\frac8N+\frac{18}{N^2}\right ) \, \tr {\phi _{0}}^2  \, \tr [\phi _{0},\phi _{1}]\phi_{1}{\phi _{0}}^4\phi
 _{1}$
 & \\[2mm]

 & & \hfill$+ \left ( 1\pm \frac4N\right )  \, \left ( 1\pm \frac6N\right )  \,
 \tr{\phi _{0}}^2  \, \tr[\phi _{1}, \phi _{0}]{\phi _{0}}^2{\phi _{1}}^2{\phi _{0}}^2
 \pm \left ( 1\pm \frac2N\right )  \, \left ( 1\pm \frac3N \right )
 \, \tr [\phi _{0},\phi _{1}]\phi _{1}{\phi _{0}}^6\phi _{1}$

 & \\[2mm]

 & & \hfill$\pm \left ( 1\pm \frac3N\right )  \, \left ( 1\pm \frac6N\right )
   \,\tr\rnode{A}{\phi _{1}}\phi _{0}\rnode{B}{\phi _{0}}\phi _{1}{\phi _{0}}^5\phi _{1}
 \pm\left ( 1\pm \frac5N\right )  \, \left ( 1\pm \frac6N\right )  \, \tr [\phi _{0},\phi _{1}]{\phi _{0}}^3{\phi_{1}}^2{\phi
 _{0}}^3$
\ncbar[linewidth=.01,nodesep=2pt,arm=.15,angle=-90]{-}{A}{B}

 &  \\[2mm]\hline
\end{tabular}
\caption{$\alSU(2)$ eigenstates from Table \ref{table:t2su2}.\label{table:t4su2}}
\end{sidewaystable}



\providecommand{\href}[2]{#2}\begingroup\raggedright\endgroup

\end{document}